\documentstyle[referee]{mn}

\newif\ifAMStwofonts

\ifoldfss
  \ifCUPmtlplainloaded \else
    \NewTextAlphabet{textbfit} {cmbxti10} {}
    \NewTextAlphabet{textbfss} {cmssbx10} {}
    \NewMathAlphabet{mathbfit} {cmbxti10} {} 
    \NewMathAlphabet{mathbfss} {cmssbx10} {} 
  \fi
  \ifAMStwofonts
    \ifCUPmtlplainloaded \else
      \NewSymbolFont{upmath} {eurm10}
      \NewSymbolFont{AMSa} {msam10}
      \NewMathSymbol{\upi}     {0}{upmath}{19}
      \NewMathSymbol{\umu}     {0}{upmath}{16}
      \NewMathSymbol{\upartial}{0}{upmath}{40}
      \NewMathSymbol{\leqslant}{3}{AMSa}{36}
      \NewMathSymbol{\geqslant}{3}{AMSa}{3E}

    \fi
  \fi
\fi 

\ifnfssone
  \newmathalphabet{\mathit}
  \addtoversion{normal}{\mathit}{cmr}{m}{it}
  \addtoversion{bold}{\mathit}{cmr}{bx}{it}
  \newmathalphabet{\mathbfit} 
  \addtoversion{normal}{\mathbfit}{cmr}{bx}{it}
  \addtoversion{bold}{\mathbfit}{cmr}{bx}{it}
  \newmathalphabet{\mathbfss} 
  \addtoversion{normal}{\mathbfss}{cmss}{bx}{n}
  \addtoversion{bold}{\mathbfss}{cmss}{bx}{n}
  \ifAMStwofonts
    \ifCUPmtlplainloaded \else
      \UseAMStwoboldmath
      \makeatletter
      \new@mathgroup\upmath@group
      \define@mathgroup\mv@normal\upmath@group{eur}{m}{n}
      \define@mathgroup\mv@bold\upmath@group{eur}{b}{n}
      \edef\UPM{\hexnumber\upmath@group}
      \new@mathgroup\amsa@group
      \define@mathgroup\mv@normal\amsa@group{msa}{m}{n}
      \define@mathgroup\mv@bold\amsa@group{msa}{m}{n}
      \edef\AMSa{\hexnumber\amsa@group}
      \makeatother
      \mathchardef\upi="0\UPM19
      \mathchardef\umu="0\UPM16
      \mathchardef\upartial="0\UPM40
      \mathchardef\leqslant="3\AMSa36
      \mathchardef\geqslant="3\AMSa3E
    \fi
  \fi
\fi 

\ifnfsstwo
  \DeclareMathAlphabet{\mathbfit}{OT1}{cmr}{bx}{it}
  \SetMathAlphabet\mathbfit{bold}{OT1}{cmr}{bx}{it}
  \DeclareMathAlphabet{\mathbfss}{OT1}{cmss}{bx}{n}
  \SetMathAlphabet\mathbfss{bold}{OT1}{cmss}{bx}{n}
  \ifAMStwofonts
    \ifCUPmtlplainloaded \else
      \DeclareSymbolFont{UPM}{U}{eur}{m}{n}
      \SetSymbolFont{UPM}{bold}{U}{eur}{b}{n}
      \DeclareSymbolFont{AMSa}{U}{msa}{m}{n}
      \DeclareMathSymbol{\upi}{0}{UPM}{"19}
      \DeclareMathSymbol{\umu}{0}{UPM}{"16}
      \DeclareMathSymbol{\upartial}{0}{UPM}{"40}
      \DeclareMathSymbol{\leqslant}{3}{AMSa}{"36}
      \DeclareMathSymbol{\geqslant}{3}{AMSa}{"3E}
    \fi
  \fi
\fi 

\ifCUPmtlplainloaded \else
  \ifAMStwofonts \else 
    \def\upi{\pi}
    \def\umu{\mu}
    \def\upartial{\partial}
  \fi
\fi

\def\simlt{\mathrel{\rlap{\lower 3pt\hbox{$\sim$}}
        \raise 2.0pt\hbox{$<$}}}
\def\simgt{\mathrel{\rlap{\lower 3pt\hbox{$\sim$}}
        \raise 2.0pt\hbox{$>$}}}

\title{On the Misalignment of Jets in Microquasars}
\author[Thomas J. Maccarone]
       {Thomas J. Maccarone \\ 
Astrophysics Sector, Scuola Internazionale Superiore di Studi Avanzati, via Beirut, n. 2-4, Trieste, Italy, 34014}

\date{}

\pagerange{\pageref{firstpage}--\pageref{lastpage}}
\pubyear{}

\begin{document}

\maketitle
\label{firstpage}
\input epsf
\begin{abstract}
We discuss the timescales for alignment of black hole and accretion
disc spins in the context of binary systems.  We show that for black
holes that are formed with substantial angular momentum, the alignment
timescales are likely to be at least a substantial fraction of the
systems' lifetimes.  This result explains the observed misalignment of
the disc and the jet in the microquasar GRO J 1655-40 and in SAX J
1819-2525 as being likely due to the Bardeen-Petterson effect.  We
discuss the implications of these results on the mass estimate for GRS
1915+105, which has assumed the jet is perpendicular to the orbital
plane of the system and may hence be an underestimate.  We show that
the timescales for the spin alignment in Cygnus X-3 are consistent
with the likely misalignment of disc and jet in that system, and that
this is suggested by the observational data.

\end{abstract}


\section{Introduction}

It is often stated that astrophysical black holes present one of the
best laboratories available for the study of general relativity.  At
the same time, relatively few concrete tests of general relativity
have been made with observations of accreting black holes.  The
Lense-Thirring effect is the distortion of space in the presence of a
rotating compact object through the dragging of inertial frames.  When
the effect is important, the central part of an accretion disc is
forced to rotate in the same plane as the black hole (the
Bardeen-Petterson effect - see Bardeen \& Petterson 1975).  The
geometry of the accretion flow changes significantly in the vicinity
of the black hole and hence to allow for sensitive tests of general
relativity.  The radius out to which this effect is observed is such
that the Lense-Thirring precession period due to the warping of the
disc is equal to the timescale for angular momentum to drift through
the disc.  Past claims of evidence for the BP effect have come from
interpretations of the quasi-periodic oscillations in X-ray binaries
(Stella \& Vietri 1998; see also Fragile, Mathews \& Wilson 2001 for a
more recent discussion).

The jets of radio galaxies are often seen to show a constant spatial
direction over their $10^8$ year lifetimes (Alexander \& Leahy 1997;
Liu, Pooley, \& Riley 1992).  Until recently, the origin for this
stability had been suggested to be that the spin of the black hole
provides a ``flywheel'' effect.  Assuming that the jets are powered by
disc accretion, the jets should be perpendicular to the discs.  The BP
effect requires that the disc plane near the black hole be
perpendicular to the black hole's spin vector.  Given the suggestion
that relativistic jets may be powered by extracting the spin energy of
the central black hole (Blandford \& Znajek 1977) and the expectation
that the timescale for the black hole to align itself with the disc by
accreting matter with net angular momentum should be at least $10^8$
years (Rees 1978), this picture became commonly accepted.

More recent observational and theoretical work challenged this picture
for the case of active galactic nuclei.  The inner dust discs of some
samples of nearby Seyfert galaxies are found to be perpendicular to
their radio jets (van Dokkum \& Franx 1995; but see Kinney et al. 2000
for contradictory results).  The jets in radio galaxies, on the other
hand, appear not to be perpendicular to the dust discs (Schmitt et
al. 2002).  If the Bardeen-Petterson effect is important and the black
hole spins are not intrinsically correlated with the dust discs'
angular momentum vectors, then alignment should be observed only in
the case of the initial alignment of black hole and disc spins, as the
size scales on which HST can resolve the discs are substantially
larger than the BP radius.  The observations of alignments in the
early work on Seyferts prompted theorists to explain the steady
directions of AGN jets by making the accretion disc, rather than the
black hole the ``flywheel'' (Natarajan \& Pringle 1998) and showing
that the black hole's spin should be much more quickly aligned with
the disc's angular momentum if one took into account the efficient
transfer of angular momentum in the disc plane (Papaloizou \& Pringle
1983) as shear perpendicular to the disk place can be transported more
efficiently than shear within the disk plane, so warp in a nonplanar
disc may decay on a timescale much faster than the accretion timescale
for a planar disk.  Additional theoretical support for the basic
assumptions in Natarajan \& Pringle (1998) has come from more detailed
calculations, both analytical (Ogilvie 1999) and numerical (Torkelsson
et al. 2000; Gammie, Goodman \& Ogilvie 2000) but see also Nelson \&
Papaloizou (2000) who suggest a longer timescale; all of these papers
note that the exact timescales depend on details of angular momentum
transport which are currently poorly understood.  The angular momentum
evolution of black holes in binary systems was studied by King \& Kolb
(1999), where it was found that black hole spins are unlikely to
change substantially due to accretion, but has not been re-visited
given the hypothesis of efficient warp transfer suggested by
Papaloizou \& Pringle (1983), and the specific problem of the change
in the spin direction of a rapidly rotating black hole has not been
considered.

In two X-ray binaries, GRO J 1655-40 and SAX J 1819-2525 (also known
as V4641 Sag), measured binary orbital plane inclination angles differ
substantially from the measured jet inclination angles, providing
strong evidence that the Bardeen-Petterson effect may be relevant.  A
third system, SS 433, shows precessing jets (Hjellming \& Johnston
1981) which imply that either the jets are not perpendicular to the
orbital plane, or that the orbital plane itself precesses, which could
occur if the system is a hierarchical triple (e.g. Fabian et
al. 1986).  With this motivation, we compute the timescale for a black
hole in a binary system to align its spin with the angular momentum of
the binary orbit, and show that in contrast to active galactic nuclei,
X-ray binaries containing black holes should routinely have misaligned
jets.  We consider the implications for the mass estimate of GRS
1915+105, which assumes that the jet is perpendicular to the binary
plane.  We discuss the case of Cygnus X-3, whose orbital inclination
angle is not well constrained, and show that its jet would have become
aligned with the orbital plane only for an unlikely range of
parameters.  We do not apply this model to SS 433, because the nature
of its binary companion is not well known and it is often speculated
that this system is undergoing thermally unstable, highly
super-Eddington mass transfer (e.g. Verbunt \& van den Heuvel 1995 and
references within), for which our thin disc approximations are highly
invalid.

\section{Applications to Binary Systems}

Assuming a thin disc model (Shakura \& Sunyaev 1973; Collin-Souffrin
\& Dumont 1990), NP calculated the timescale for alignment of the
black hole's angular momentum vector with that of the accretion disc:

\begin{equation}
t_{align} = 5.6\times10^5 a^{11/16}
(\frac{\alpha}{0.03})^{13/8}(\frac{L}{0.1L_E})^{-7/8}M_8^{-1/16}(\frac{\epsilon}{0.3})^{7/8} \rm{years},
\end{equation} 
where $a$ is the dimensionless angular momentum parameter for the
black hole, $\alpha$ is the viscosity parameter of the accretion disc
in the Shakura-Sunyaev (1973) formalism, $L$ is the steady luminosity
of the accretion flow, $L_{E}$ is the Eddington luminosity for the
central black hole, $M_8$ is the black hole mass in units of $10^8$
solar masses, and $\epsilon$ is the efficiency of the accretion flow,
i.e. $L/\dot{m}c^2$, where $\dot{m}$ is the accretion rate.  This
timescale represents a decay timescale for the separation between the
disc's angular momentum axis and the jet's axis, and if a system's age
is less than a few alignment timescales, some observable separation
between disc and jet axes is to be expected if the initial separation
was large.

We re-write this in terms of the accretion timescale,
$t_{acc}=M/\dot{m}$ and set the fiducial value of the black hole mass
to 10 solar masses, a more typical value for binary systems.

\begin{equation}
\frac{t_{align}}{t_{acc}} = 1.3\times10^{-2} a^{11/16} (\frac{\alpha}{0.03})^{13/8}(\frac{L}{0.1L_E})^{1/8}M_{10}^{-1/16}(\frac{\epsilon}{0.3})^{-1/8}
\end{equation} 

The lifetimes, $\tau_{bin}$ of Roche lobe overflow binary systems have
as an upper bound the timescale to accrete the entire donor star -
$\frac{M_{don}}{M_{CO}}t_{acc}$, where $M_{don}$ is the mass of
the donor star and $M_{CO}$ is the mass of the accreting compact
object.  Making this substitution, we find:

\begin{equation}
\frac{t_{align}}{\tau_{bin}} = 0.13 a^{11/16}(\frac{\alpha}{0.03})^{13/8}(\frac{L}{0.1L_E})^{1/8}M_{10}^{15/16}(\frac{\epsilon}{0.3})^{-1/8} (\frac{M_{don}}{M_\odot})^{-1}. 
\end{equation} 
In fact, the lifetimes of binary systems may be quite a bit shorter
than $\tau_{bin}$.  Winds from the donor star, perhaps driven by X-ray
heating from the accretion flow, may accelerate the mass loss of the
donor star without changing the accretion rate on the compact object.
Such a scenario has been suggested for reducing the lifetimes of low
mass X-ray binaries containing neutron stars in order to match the
LMXB birth rate with the millisecond pulsar birth rate (e.g. Tavani
1991; Podsiadlowski 1991) .  

A substantial fraction of black hole binaries in the Galaxy are
transient systems.  These induce significant complications in applying
the above formulae.  First, their time averaged luminosities are not
always well known, as the recurrence timescales for the outbursts are
poorly constrained for most systems.  Secondly, the transfer of
angular momentum of a warp is enhanced in large scale height flows.
Since a prominent model for the quiescent states of accreting binaries
is the advection dominated accretion flow model, where the scale
height goes roughly as the radius, the above formulae become invalid.

Two other issues make the problems related to geometrically thick
flows less worrisome.  First, despite the fact that transient systems
spend most of their time in the quiescent state, they emit the vast
majority of their luminosity in the outbursts.  If we assume the
relation of Narayan \& Yi (1995) that $\epsilon\sim
\dot{m}/\dot{m}_{EDD}$, a $\sim$ 6 year recurrence timescale, a 30\%
efficiency in outburst, and that the outbursts consist of emission at
$\sim L_{EDD}$ for a few months, while the quiescent periods consists
of emission at $10^{-7} L_{EDD}$, then only a few percent of the
matter is accreted during the quiescent states.  Thus even if the
accretion flow is 10 times more efficient at transferring warps in
quiescence, the transferring of warps during the outbursts dominates.
The second issue is that much recent theoretical work has suggested
that outflows from the accretion flow may be extremely important in
the quiescent states (Blandford \& Begelman 1999; Quataert \& Narayan
1999; di Matteo et al. 1999; Meyer, Liu \& Meyer-Hofmeister 2001).  If
matter is expelled from the accretion flow because it retains excess
angular momentum, then obviously, the efficiency of this matter in
transferring angular momentum from the disc to the black hole is
reduced.

Thus it is safe to assume that for transient black hole systems, the
accretion history can be approximated as a series of discrete events
where the system accretes at $\epsilon \sim30$\% efficiency and $L
\sim L_{EDD}$.  The expressions then become:

\begin{equation}
\frac{t_{align}}{\tau_{bin}} = 0.10 a^{11/16}(\frac{\alpha}{0.03})^{13/8}(\frac{L_{out}}{L_E})^{1/8}M_{10}^{15/16}(\frac{\epsilon}{0.3})^{-1/8} (\frac{M_{don}}{M_\odot})^{-1}.
\end{equation}

For systems with low mean accretion rates, the Hubble time and/or the
stellar evolution timescale for the companion star may be shorter than
the binary lifetime and/or the alignment timescale.  In particular,
for high mass X-ray binaries (i.e. wind-fed accreting systems), the
system lifetime will almost always be determined by the stellar
evolution timescale for the donor star.  Many transient systems with
low mean accretion rates will also have donor stars with lifetimes
shorter than $\tau_{bin}$.  Thus we wish to know the alignment
timescale in years as well as in terms of $\tau_{bin}$, so we can
compare with the Hubble time and with the main sequence lifetime of
the companion star.  Expressing $\tau_{bin}$ explicitly, we find:

\begin{equation}
\tau_{bin} = 1.3 \times 10^8 (\frac{M_{don}}{M_{CO}})
(\frac{\epsilon}{0.3}) (\frac{L_{out}}{L_{EDD}})^{-1}
(\frac{t_{recur}}{t_{out}}) \rm{years}.
\end{equation}
Thus for binaries which spend a small fraction of their time in
outburst, and hence have low mean mass accretion rates, $\tau_{bin}$
can be of order the Hubble timescale.  The alignment timescale can
then be expressed as:

\begin{equation}
t_{align} = 1.0\times 10^7
a^{11/16}(\frac{\alpha}{0.03})^{13/8}(\frac{L_{out}}{L_E})^{-7/8}M_{10}^{-1/16}(\frac{\epsilon}{0.3})^{7/8} (\frac{t_{recur}}{t_{out}}) \rm{years}.
\end{equation}

\section{Implications for the Misalignment of the Jet in GRO J 1655-40 and SAX 1819-2525}

The first known and one of the strongest candidates for the
Bardeen-Petterson effect is GRO J 1655-40.  Its misalignment between
binary plane (70 degrees - Greene, Bailyn \& Orosz 2001) and jet (85
degrees - Hjellming \& Rupen 1995) of at least 15 degrees has been
suggested to be evidence for the Bardeen-Petterson effect (Fragile,
Matthews \& Wilson 2001).  Based on the observations of 300 Hz and 450
Hz quasi-periodic oscillations in this source (Strohmayer 2001a),
thought to be in a resonance, its value of $a$ has been estimated to
be between 0.2 and 0.67 (Abramowicz \& Kluzniak 2001), with most of
the uncertainty from the uncertainty in the mass estimate of the black
hole.  Alternatively, if the 300 Hz QPO represents a diskoseismic
$g$-mode, and the 450 Hz represents its corresponding $c$-mode
(Wagoner, Silbergleit, \& Ortega-Rodriguez 2001 - WSO), then
$a=0.92\pm0.02$.  Taking the relatively conservative case of 6.3 solar
mass black hole, a spin parameter value of 0.4, $\alpha=0.05,
\epsilon=0.3$, a companion star mass of 2.4 solar masses and
$L=L_{EDD}$ in outburst, we find that the ratio of alignment timescale
to binary lifetime is about 0.3, without invoking any winds, and with
the binary lifetime assumed to be the timescale to accrete the whole
companion star, not the stellar evolution timescale of the companion
star.

We now check the lifetime in years rather than the ratio of the two
timescales to test whether stellar evolution may place a more
stringent limit on the lifetime of the system than accretion does.
The outburst history of GRO J 1655-40 is somewhat complicated - it has
shown several outbursts over the last 10 years, but previously seems
to have been quiescent for quite some time.  As a southern hemisphere
source, it is not well observed in the optical plate archives often
used to find past outbursts of novae.  The typical recurrence
timescale is thus rather uncertain, so we take the mean values for
outburst duration and recurrence timescales for transient objects
(e.g. Chen, Shrader, \& Livio 1997 - CSL).  The recurrence timescale
is the assumed to be about 6 years.  The outburst durations for
observability are seen to be about 100 days by CSL, but the peaks
during which $L\sim$$L_{EDD}$ are somewhat shorter.  We thus take a 30
day recurrence timescale.  This gives us an alignment timescale of
about $8\times10^{8}$ years.

Assuming that a star's main sequence lifetime is $\sim 10^{10}
(M/M_{\odot})^{-3}$ years for stars below 30 solar masses
(e.g. Kippenhahn \& Weigert 1990), then the main sequence lifetime of
the 2.4 $M_{\odot}$ secondary star in GRO J 1655-40 should be about
$7\times10^{8}$ years.  The secondary appears to be somewhat evolved
(Greene, Bailyn, \& Orosz 2001), so the age of the system is likely to
be a substantial fraction of the star's main sequence lifetime.  Thus
for our best guess regarding the recurrence timescale (and hence the
mean luminosity), we find that the system should be $\sim$ one
alignment timescale in age, and hence the black hole's spin should not
have become fully aligned with that of the accretion disc.  If the
diskoseismic model, rather than the epicycle-resonance model, is
correct, the spin parameter value will be a factor of $\sim$2.5 higher
and the alignment timescale will then be about twice the system's
age. It should be noted that the age estimate for the secondary star
is an upper limit; if the star has undergone significant mass loss
through stellar winds, then its initial mass was larger than the
current measurement shows and its stellar evolution timescale will be
shorter than our estimate.  We caution alternatively that the value of
$\alpha$ could be substantially smaller than 0.05 and that the
dependence of alignment timescale on $\alpha$ is rather strong.  In
this case, the system could be misaligned despite have an age of more
than one alignment timescale.  Such a result could be explained by
outflows and/or jets taking away a substantial fraction of the
system's angular momentum, and could have important implications for
the observed misalignments of AGNs which are several alignment
timescales old.  Figure 1 shows the three timescales - the stellar
evolution timescale, the timescale to accrete the donor star, and the
alignment timescale as a function of the duty cycle.

\begin{figure*}
\centerline{\epsfxsize=12.4 cm \epsfysize=10.6cm \epsfbox{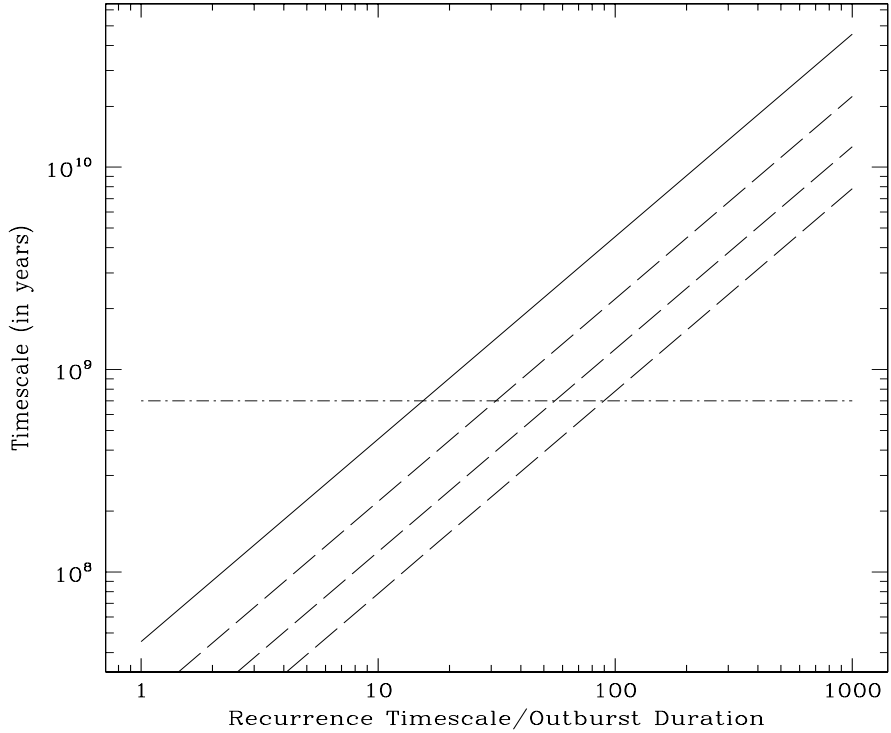}}
\caption {The three relevant timescales, as a function of the ratio of
recurrence timescale to outburst timescale.  The solid line is the
timescale to accrete the entire donor star.  The dot-dashed line is
the main sequence stellar lifetime of the 2.4$M_\odot$ donor star.
The three dashed lines are the alignment timescales for spins of 0.92,
0.4, and 0.2, from top to bottom.  The other parameters are chosen as
specified in the discussion of the parameter values for GRO J
1655-40.}
\end{figure*}

A similar line of reasoning shows that if the black hole in SAX
1819-2525 (V4641 Sag) was formed with a substantial spin, it, too
should spend the whole binary lifetime unaligned.  Orosz et al. (2001)
find that the mass donating star in this system is a late B type star
of about 6 $M_\odot$, meaning that its main sequence lifetime is of
order $10^8$ years or less (as the star has likely undergone
significant mass loss).  They find a 10 solar mass black hole and a
distance of about 10 kpc.  The jet's spatial extent within one day of
the start of the X-ray outburst then implies an apparent velocity of
$\sim10c$ (Hjellming et al. 2001).  Thus the jet must be strongly
beamed (with the Lorentz factor of the jet projected along the line of
sight, $\delta, \sim 10$), implying an inclination angle of less than
about 10 degrees (Orosz et al. 2001), while measurements of the May
2002 outburst show even larger proper motions, implying $\delta\sim17$
and hence an inclination angle less than 7 degrees (M. Rupen, private
communication).  Orosz et al. (2001) also find strong ellipsoidal
variations, indicating a binary plane inclination angle of at least
$\sim 60$ degrees.  The angle between the orbital plane and the jet
must then be at least $\sim50$ degrees. Since the system appears to be
a transient that even in outburst emits at a small fraction of its
Eddington luminosity (with the exception of the brief, beamed jet
episode - see Wijands and van der Klis 2000), its alignment timescale
should be well in excess of $10^8$ years unless it was formed with a
very low value of the spin parameter $a$ (one less than $10^{-3}$,
assuming that the mean luminosity of V4641 Sag is ten times lower than
that of GRO J 1655-40).  If the black hole spin affects the production
of jets (as suggested, for example, by Blandford \& Znajek 1977), we
would then not see jets from a source with such a low spin.

\section{Implications for Mass Estimates in GRS 1915+105}

For GRS 1915+105, the disc and jet have often been assumed to be
perpendicular, both for purpose of computing the binary inclination
(Greiner, Cuby, \& McCaughrean 2001 - GCM) and for purposes of
computing the inner disc radii given a normalization to the disc
component of the spectrum (e.g. Muno, Remillard \& Morgan 1999;
Klein-Wolt et al. 2001).  The companion star's mass is estimated to be
about 1.2 $M_{\odot}$ and the black hole's mass is estimated to be
about $14\pm4$ solar masses.  The inclination angle given from jet
measurements is 70 degrees (Mirabel \& Rodriguez 1994), so the black
hole mass cannot have been overestimated by much more than 5\% if the
inclination angle of the binary plane is not the direction
perpendicular to the jets.  The black hole mass could, however, be
severely {\it underestimated} if the jet is not perpendicular to the
accretion disc.

We thus consider other estimates of the black hole mass and attempt to
determine whether the spin vectors of the orbit and the black hole
should be aligned given these masses and the expected values of the
spin parameter.  WSO present two estimates for the mass of GRS
1915+105 from diskoseismology based on the measurements of two high
frequency QPOs (Strohmayer 2001b).  The first, based on the
identification of the 67 Hz QPO with the $g$-mode, as suggested by
Nowak et al. (1997), gives a mass of $18.2\pm3.1$ solar masses and
$a=0.70\pm0.04$.  The 42 Hz QPO is then the $c$-mode.  The second
estimate consists of switching the identifications.  If the more rapid
QPO is the $c$-mode, as seen in GRO J 1655-40, then the mass of the
black hole is $42.4\pm7.0$ solar masses and $a=0.93\pm0.02$.  In
either case, the spin alignment timescale for the black hole should be
substantially longer than the lifetime of the system.  The 18 solar
mass estimate is within the error bars of the measurement of GCM,
while the 42 solar mass estimate would require the inclination angle
of the binary plane to be about 20 degrees, implying a misalignment of
disc and jet of at least 50 degrees.  WSO state that the higher mass
estimate is consistent with the findings of Strohmayer (2001b) that
the relative spectral properties of high and low frequency QPOs in the
two microquasars systems are the same.  

More recent measurements of additional high frequency QPOs in GRS
1915+105 at 162 and 324 Hz, with an additional possible detection at
486 Hz (Remillard et al. 2002, in preparation) cast serious doubt on
the diskoseismic model.  A consistent set of values for the mass and
spin cannot reproduce both the 67 and 40 Hz QPOs as diskoseismic modes
and the higher frequency QPOs as a 1:2 or 1:3 resonance between
orbital and epicyclic frequencies.  The resonance model rules out a
large misalignment between the disc and jet inclination angles.
Figure 2 shows the curves for the 1:2 and 1:3 resonances which give an
epicyclic frequency of 162 Hz (with orbital frequencies of 324 and 486
Hz respectively), along with the curve where the disc and jet are
misaligned.  We compute the relation between the minimum spin and mass
needed for a misaligned system by assuming $L=3\times10^{38}$ ergs/sec
(the mean value as estimated from the RXTE All Sky Monitor data), a
30\% efficiency, $\alpha$=0.05. Since the initial mass of the
companion star is not known, we take the as an upper limit the mass
where the time to accrete all but 1.2 $M_\odot$ at 30\% efficiency at
the current mean luminosity equals the stellar evolution timescale for
a star at that mass.  For these parameter values, the maximum initial
mass of GRS 1915+105's companion is about 3.6 $M_\odot$.  The system
has then lived at most $1.5 (M/14 M_\odot)^{13/16} a^{11/16}$
alignment timescales, under the assumptions that all the matter
accreted makes it into the hole, that mass loss by the companion star
is unimportant, and that the system has been in outburst for its
entire lifetime.  The third assumption here is known to be false and
makes the estimate an upper limit.  If we require three alignment
timescales for full alignment, then the system may be aligned only for
$a<0.36 (M/14 M_\odot)^{-13/11}$, bearing in mind that this is a
conservative estimate since we know that the luminosity of the system
was much lower before its discovery in 1992 than it has been during
the RXTE era.  We plot in Figure 2 three curves - two showing the loci
of points allowed by the 1:2 and 1:3 resonance models for the QPOs and
the third showing the loci of points where the black hole has had time
to align its spin with that of the binary plane.  From inspection of
the curves, one can see that the strong upper bound on the mass of GRS
1915+105 comes from the QPO frequencies, and not from the measurement
of the orbital parameters, as the alignment timescale is substantially
longer than the expected lifetime of the system.  The inclination
angle of the orbital plane of GRS 1915+105 cannot be more than $\sim
25$ degrees offset from the jet angle if the resonance model of the
quasi-periodic oscillations is correct.  Thus while the mass of GRS
1915+105 itself cannot be too seriously in error due to disc-jet
misalignment, this system should serve as a cautionary tale when
future mass measurements are made assuming that the disc and jet are
aligned.

\begin{figure*}
\centerline{\epsfxsize=12.4 cm \epsfysize=10.6cm \epsfbox{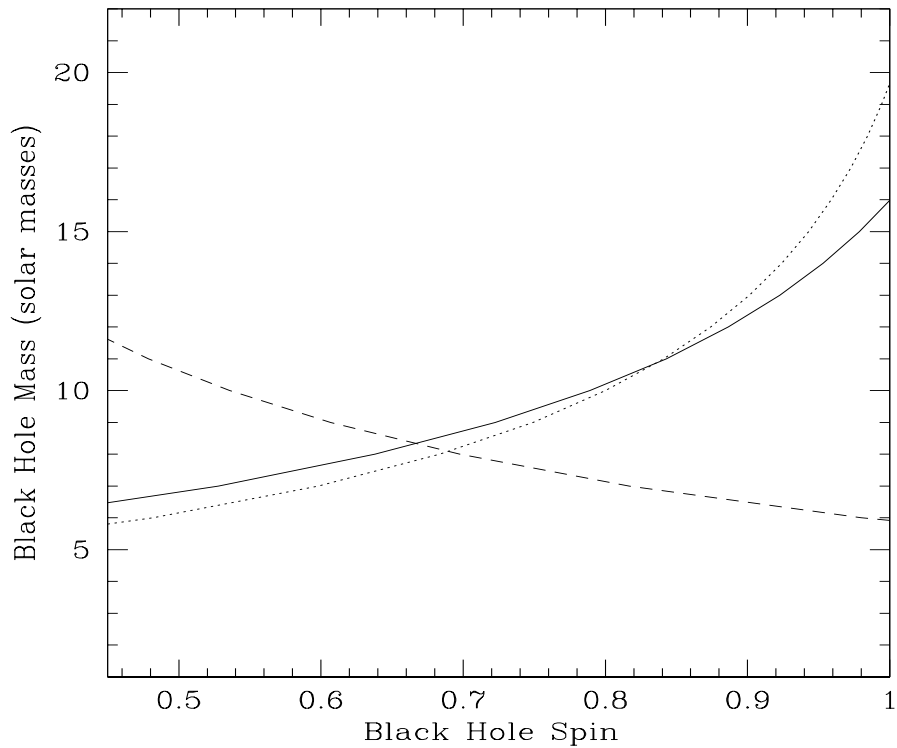}}
\caption {The spin-mass curves for GRS 1915+105.  The solid curve
represents the locus of points which provide a 162 Hz epicyclic
frequency in resonance with a 324 Hz orbital frequency.  The dotted
line provides a 162 Hz epicyclic frequency in resonance with a 486 Hz
orbital frequency.  The dashed line shows where the timescale for
alignment of the black hole's spin equals its maximum age as
calculated in th text.  Above the line, the inclination angle of the
system can deviate substantially from the jet inclination angle, and
the mass measurement based on the orbital parameters may be
subtantially in error.}
\end{figure*}

We also note that misalignment of disc and jet in GRS 1915+105 has
already been speculated on the basis of the enhanced Fe and Si
abundances in the system's absorption column (Lee et al. 2002).  In
their scenario, polar ejecta (which are richer in heavy elements than
the non-polar ejecta) from a supernova/hypernova that produced the
black hole in GRS 1915+105 may have contaminated the atmosphere of the
companion star.  While we find that a jet inclined in the binary plane
of the system is inconsistent with the allowed parameter space (due to
the maximum black hole mass allowed by the QPO measurements), the fact
that the companion star occupies a rather small fraction of the binary
plane (as its radius is only about 1/10 of the binary separation) and
that the orbital period is longer than the ejection phase implies that
a rather large opening angle for the polar ejection would be required
for this scenario to work anyways.  Furthermore, if the misalignment
of the jet is in the position angle, rather than in the inclination
angle, the jet could still be parallel to the plane of the binary
orbit.  At present there are no constraints on the position angles of
the binary orbits, but constraints for some systems could come in the
next decade from SIM and/or MAXIM.

\section{Application to Cyg X-3}

Cygnus X-3 is a persistent emitter in the X-rays and in the radio,
with flaring episodes where it is one of the brightest sources in the
sky in both those wavebands.  It has been recently found to have a
one-sided radio jet with an inclination angle of about 14 degrees
(Mioduszewski et al. 2001).  That only one side of the jet is seen
suggests that the velocity of the jet is at least 0.8$c$.  Like GRS
1915+105, its optical and infrared properties are rather poorly
constrained because of the high column density to the source and
because of the contaminating emission from the accretion flow.  Cyg
X-3 is known to be a high mass X-ray binary accreting from the stellar
wind of a Wolf-Rayet star of somewhere between 5 and 20 solar masses.
A 4.8 hour period has been seen in both the infrared and X-ray
emission and is assumed to be the orbital period.  The physically
simplest and statistically best fitting model for the orbital
parameters - that of a constant strength wind and an elliptical orbit,
requires an inclination angle of $51\pm2$ degrees (Ghosh et al. 1981).
In this case, there is at least a 37 degree angle between the jet and
the direction perpendicular to the binary plane.  Hanson et al. (2000)
also find that the system is unlikely to be at an inclination angle as
low as the 14 degrees of the radio jet.  Attempts to measure the
orbital plane through polarimetry have been unsuccessful, probably
because the jet emission contaminates the infrared light curve (Jones
et al., 1994).  Using the 51 degree inclination angle, the estimated
mass of the compact object is 17 solar masses (Schmutz et al. 1996),
and in fact, if the inclination angle of the binary system {\it is} 14
degrees, the mass of the compact object would be about 50 $M_\odot$.
Given a typical luminosity of approximately 5\% of Eddington (as
estimated from the RXTE All Sky Monitor public data) and a stellar
lifetime for the presumed 20 solar mass progenitor of the Wolf-Rayet
companion star, we estimate that the initial spin of the black hole
would have to have been only about $a=5\times10^{-4}$ in order not to
have been aligned.  Even given the much smaller mass estimate of
Hanson et al. (2000), which states that the system may contain a much
less massive black hole, $a$ must be less than about $10^{-3}$ for the
black hole's spin to have aligned with that of the binary plane.

The compact star in Cyg X-3 may be a neutron star (Hanson et
al. 2000).  Since the angular momentum of a maximally rotating neutron
star is about the same as that for a maximally rotating black hole of
1.4 $M_\odot$, we use the same formula as for the black hole case to
estimate that the alignment timescale will be about $10^7$ years,
which is substantially longer than the estimated $10^6$ year lifetime
of the companion star.  Thus there is no strong constraint on the
inclination angle of the binary plane from the jet inclination angle,
and a neutron star primary, which would require $i>60$ degrees, is
still permitted by the data.  The only way to force alignment is if
the characteristic spin-down timescale for the neutron star is
sufficiently fast that its initial angular momentum when accretion
begins is substantially smaller than that of a maximally rotating
neutron star.  The spin-down timescale is $10^6$ years for a magnetic
field of $4\times10^{12}$ Gauss, after which the pulsar period is of
order 1 second.  This is also roughly the equilibrium period for a
$4\times10^{12}$ Gauss neutron star accreting near the Eddington limit
(e.g. Bhattacharya 1995).  Thus a only a very high magnetic field
neutron star can be ruled out, as its inner disc would be aligned with
the neutron star's spin axis.  High magnetic field neutron stars
accreting in high mass X-ray binaries are generally observed to be
accretion powered pulsars, and coherent pulsations have not been seen
in Cyg X-3, so this system is unlikely to contain such an object.

\section{Conclusions}

We have shown that the timescale for the black hole spin to align with
the accretion disc's angular momentum in binary systems is often
longer than the lifetime of the binary system.  The Bardeen-Petterson
effect should then be important in these systems if the black holes
were formed with substantial angular momentum.  This represents a
fundamental difference between black holes in binary systems and the
black holes in active galactic nuclei which should align in a
relatively short fraction of their lifetimes (as shown by NP).  Strong
evidence of this effect is seen from GRO J 1655-40 and SAX J1819-2525.

That the timescale for spin alignment in GRO J 1655-40 is so close to
the main sequence lifetime of the companion star hints at a possible
explanation for the difference between radio galaxies (which are
unaligned) and Seyfert galaxies (which are often aligned).  The strong
jets in the radio galaxies may carry away a substantial amount of
angular momentum, keeping the matter and its angular momentum from
reaching the black hole and making the spin changing estimate of NP98
a severe underestimate.  The Seyfert galaxies have a much lower
fraction of their total power taken away by the radio jets, so this
effect will not be as severe for them.  Secondly, if the
Blandford-Znajek mechanism is responsible for powering the radio jets,
it is likely that the dimensionless spin parameter values of the black
holes in radio galaxies are higher than those in the Seyfert galaxies,
giving another reason why the timescales for the radio galaxies' black
holes to change their spins might be longer.

The dynamical mass estimate for GRS 1915+105 depends upon the
inclination angle of the system being such that the disc is
perpendicular to the jets.  We have shown that for high mass, high
spin black hole primaries in GRS 1915+105, the alignment timescale of
the black hole's spin with that of the accretion flow can be longer
than the characteristic age of the system.  On the other hand, the
orbital/epicyclic resonance model for the quasi-periodic oscillations
recently discovered in this system suggest that the black hole mass
cannot be too far outside the error bars from the dynamical
measurement.  As a result, the inclination angle separation between
the disc and the jet is likely to be less than about 25 degrees.  Thus
the system must have either been formed with a small offset between
disc and jet angle, or the offset between the two angles is largely in
the plane of the sky rather than in the inclination angle.  Future
planned X-ray interferometry missions such as MAXIM may have the
potential for measuring the position angles of systems such as GRS
1915+105 by imaging the plane of the ``hot spot'' where accretion
stream hits the accretion disc.

We discuss suggestive evidence that the disc and jet are unaligned in
Cygnus X-3, and show that this is a result to be expected in wind-fed
accreting black holes even with very small initial angular momenta.  A
neutron star origin is also consistent with the binary orbital data,
since even a rapidly rotating neutron star would have too much angular
momentum to be aligned in the lifetime of the donor star.  It is
likely that the system will need to be observed in quiescence before
polarization measurements can tell us its true orbital inclination
angle.  Future observations of inclination angles in microquasars
should help determine whether the results for these four systems are
common.

\section{Acknowledgements}

I wish to thank Priya Natarajan and Annalisa Celotti for useful
discussions and for critical reviews of this manuscript.  I also wish
to thank Arun Thampan and Wlodek Kluzniak for interesting discussions
regarding high frequency QPOs from accreting black holes.  I wish to
thank Julia Lee for discussions regarding the abundance anomolies in
GRS 1915+105.  Finally, I thank the anonymous referee for useful
suggestions.

\label{lastpage}
\end{document}